\documentclass[preprint]{elsarticle}
\pdfoutput=1 

\usepackage{amsmath}
\usepackage{caption}
\usepackage{graphicx,xcolor}      
\usepackage{longtable, gensymb}     
\usepackage{url,physics}           
\usepackage{bm}            
\usepackage[english]{babel}

\newcommand*\bob{\color{black}}
\newcommand*\glenn{\color{black}}

\begin{document}
\title{{\color{black} Death and \bob {Serious Injury from} Dark Matter}}

\author[1]{Jagjit Singh Sidhu\corref{cor1}}
\ead{jxs1325@case.edu}
\author[2]{Robert Scherrer}
\ead{robert.scherrer@vanderbilt.edu}
\author[1]{Glenn Starkman}
\ead[url]{glenn.starkman@case.edu}
\cortext[cor1]{Corresponding author}

\address[1]{Physics Department/CERCA/ISO Case Western Reserve University  \\
	Cleveland, Ohio 44106-7079, USA}
\address[2]{Department of Physics $\&$ Astronomy, Vanderbilt University,
Nashville, TN 37235}

\begin{abstract}
Macroscopic dark matter {\bob (macros)} refers to a {\bob class} of 
dark matter candidates that scatter elastically off of 
ordinary matter with
a large geometric cross-section. A wide range of macro masses $M_X$
and cross-sections $\sigma_X$ remain unprobed.
We show that over a wide region within the unexplored parameter
space,
collisions of a macro with a human body would result in serious injury or death.
We use the absence of such unexplained impacts with a well-monitored subset
of the human population to exclude a region bounded by $\sigma_X > 10^{-8} -
10^{-7}$ cm$^2$
and $M_X < 50$ kg. Our results open a new window on dark matter:  the human body as a dark
matter detector
\end{abstract}

\maketitle

\section{Introduction}

{\bob The evidence for dark matter is overwhelming (see,
e.g., \cite{a} and references therein), but}
the nature of 
dark matter remains one of the great
unsolved mysteries of modern cosmology.
Recently, Jacobs, Starkman, and Lynn \cite{b} {\color{black} explored the proposition}
that the dark matter might be macroscopic, in the sense of
having a characteristic mass $M_X$ and cross-sectional area in the gram and cm$^2$ range,
respectively.
In this model,
the macroscopic dark matter objects (dubbed ``macros") have a geometric
cross section $\sigma_X$ equal to the cross-sectional area of the macro.

{\bob Macros are most likely}
composites of more fundamental particles.
An intriguing possibility is that macros could be made
of Standard Model quarks or baryons bound by Standard Model forces. This suggestion was originally made
by Witten \cite{c}, in the context of a 
first-order QCD phase transition early in the history of the Universe. A
more realistic version was advanced by Lynn, Nelson and
Tetradis \cite{d} and Lynn \cite{e}
in the context of $SU(3)$ chiral perturbation theory.  They
argued that ``the true bound state
of nuclei may have two thirds
of the baryon number consisting
of strange quarks and that ordinary
nuclei may only be metastable." 
Nelson \cite{f} studied the possible
formation of such ``nuggets of strange baryon matter''
in an early-universe transition from a kaon-condensate phase of QCD to the ordinary phase.
Others have suggested non-Standard Model versions of such nuclear objects and their formation, for example incorporating the axion  \cite{g}.

{\bob Once the mass and cross-section
of the macros are specified, the internal density of an individual macro is completely
determined by
the fact that the cross-section is geometric.
Macros corresponding to the models mentioned in the previous paragraph would most likely have densities that are comparable to nuclear density 
(which we take to be $\rho_{nuclear}=3.6 \times 10^{14}\,$g$\,$cm$^{-3}$).
This is much higher than ordinary ``atomic density"
($\rho_{atomic}=1\,$g$\,$cm$^{-3}$), 
but much lower than the density of black holes.
Although macros of approximately nuclear density
are
of particular interest, other densities are not excluded at this point,
so we will consider the full range of possibilities for $M_X$ and $\sigma_X$.
Note that 
macros that form prior to $T \sim 1$ MeV
are not subject to the bounds on the baryon density
from Big-Bang nucleosynthesis.}

\includegraphics[width=\textwidth]{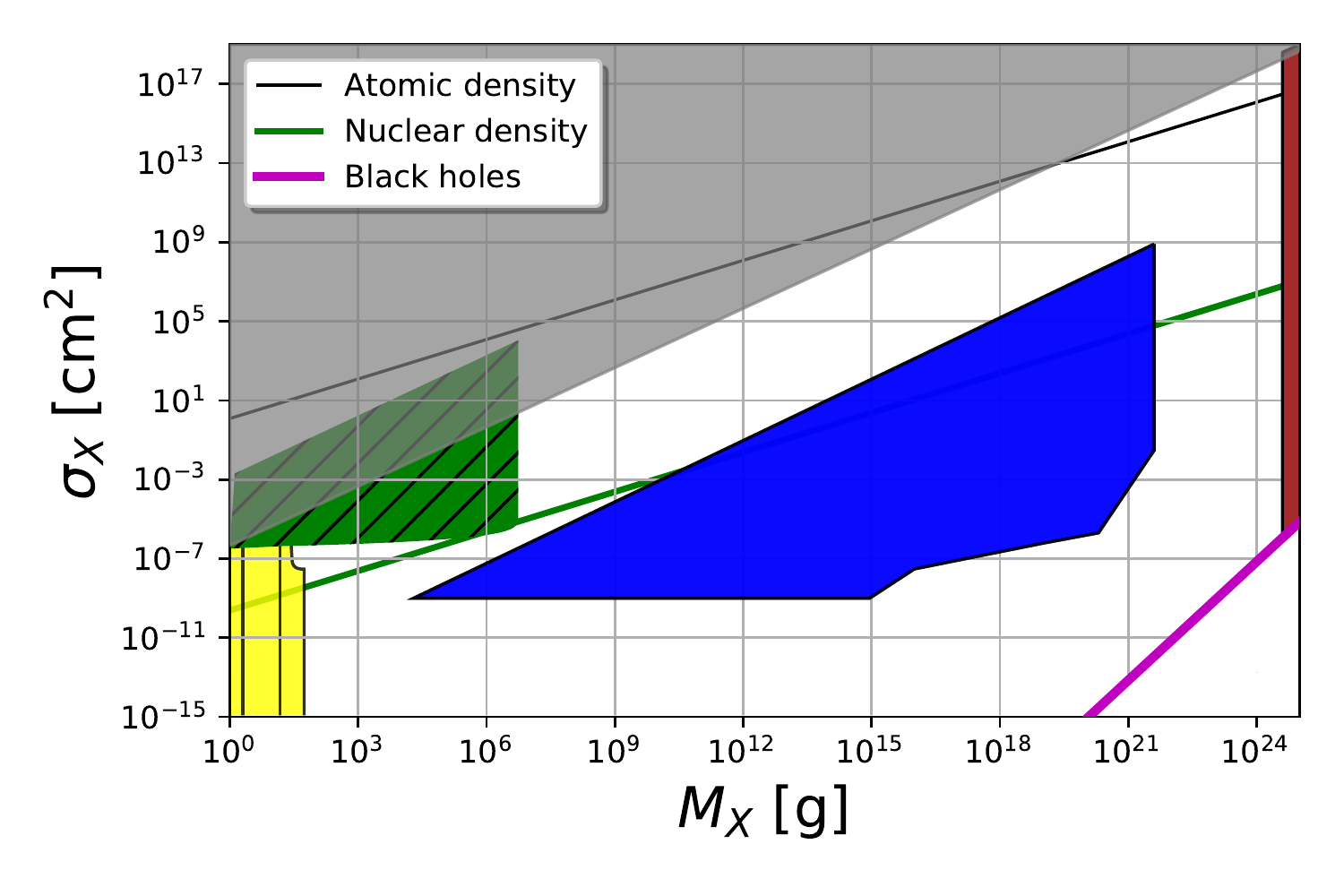}
\captionof{figure}{Constraints for macros
over a wide range of masses and 
cross-sections.
Constraints in yellow are
derived from a lack of tracks in an
ancient slab of mica \cite{h,i}, in grey 
from the Planck Cosmic Microwave Background data considering
elastic macro-photon interactions \cite{o},
in red from microlensing experiments \cite{k,l,m,n} and in blue
from thermonuclear runaway in white dwarfs \cite{o}.
We have also presented projected regions of parameter space accessible by
future searches. The region in green with hatching represents
the union of the 
region accessible using the Pierre Auger Observatory \cite{p}, 
the JEM-EUSO planned experiment\cite{p},
and a search of $\approx100$ slabs 
	of commercial granite  \cite{q}. 
The granite-slab search could be scaled up 
to access much larger macro masses (and smaller fluxes), 
e.g. through a citizen science program, 
which is the goal of two of the authors (JSS and GDS)
once a preliminary search has been completed.}

{\bob Previous work has placed a wide range of constraints on macro masses and cross sections from purely phenomenological
considerations,
which are displayed in Fig. 1}. 
For macro masses
$M_X \leq 55\,$g, careful examination of specimens of old mica
for tracks made by passing dark matter \cite{h,i} has ruled
out such objects as the primary dark-matter candidate
(see Figure 1). {\bob For very large macro masses} ($M_X \geq 10^{24}\,$g), a variety of microlensing searches have similarly constrained macros
\cite{j,k,l,m}.
A large region of parameter space was constrained by considering thermonuclear runaways triggered by
macros incident on white dwarfs \cite{n}.
{\bob For sufficiently large $\sigma_X$, scattering between photons
and macros will distort the fluctuation spectrum of the cosmic microwave background.
Reference \cite{o} utilized the first year release of Planck data to place
constraints on $\sigma_X$ and $M_X$.}

{\bob A number of other constraints have been proposed recently.}
It has been suggested that
ultra-high-energy cosmic-ray detectors that exploit
atmospheric fluoresence could be modified to
probe parts of macro parameter space \cite{p}, including
macros of nuclear density and intermediate mass.
Macros with $\sigma_x \geq 10^{-6}\,$cm$^2$
were shown to be able to produce an observable fluorescence signal
assuming changes to the time binning mechanism of 
a typical fluorescence detector.
It has also been suggested that the approach applied
to mica could be adapted to a larger, widely available sample of granite, to search for
larger-mass macros \cite{q}. 
{\bob Both these methods have the potential
to probe masses exceeding 10$^6$ g},
{\color{black} and the combined parameter space
that could be probed is highlighted in green with hatching
in Figure 1.}

{\bob In this manuscript, we consider the phenomenology
of such objects and, in particular, their effects on the human population. We
derive a new constraint on some region of the allowed macro parameter space, by
noting that for a range of macro masses and cross sections, collisions
of macros with the human population would have caused a detectable number
of serious injuries and deaths with obvious and unusual features, while
there have been no reports of such injuries and deaths in
regions of the world in which the human population is well-monitored.}  
(Previously, others \cite{r} have considered
the effects of {\bob weakly interacting massive particle} (WIMP) collisions with the human body,
{\bob with the conclusion} that WIMPs would be 
essentially harmless.)

\section{Derivation of Constraints}
{\bob Consider a macro with cross section $\sigma_X$ and velocity
$v_X$ passing through the human body.
The energy per unit length deposited by a macro 
through elastic scattering on any target is 
\begin{equation}\label{dedx}
\frac{dE}{dx} = \sigma_X \rho v_X^2,
\end{equation} 
where $\rho$ is the density of the target.
As in previous studies, we assume a sufficiently
strong interaction between macros and baryonic matter
that $\sigma_X$ is given by the geometric
cross section.
For human tissue, a
good approximation for the target density is the density of water: $\rho \sim 1\,$g cm$^{-3}$.

To determine the amount of damage produced by a macro collision, we make an analogy to
gunshot wounds (although there are significant differences, which we will discuss below).
Bullets cause injury to
the human body from a combination of permanent cavitation,
temporary cavitation, and pressure waves \cite{s}.  While
these are complex processes, it is generally believed that
the overall tissue damage depends primarily on the kinetic
energy deposited in the body.  This is the key assumption we make in this paper:  the amount
of damage caused by a macro will scale as the kinetic energy, and the damage produced will
be similar to that of a bullet that deposits a similar amount of kinetic energy.
Bullets in general
have muzzle kinetic energies in the range of 
100-10000 J \cite{t},
although only a fraction of the muzzle kinetic
energy  is deposited unless
these bullets
stop inside the body.
As our benchmark for ``significant" damage to the human body, we will take the muzzle energy (100 J)
from a .22 caliber rifle {\color{black}\cite{t,t1}.}
This is the smallest rifle in common use but is still capable of inflicting serious injury.
Hence we will require at least 100 J to be deposited by the macro
as it traverses a human body.  To determine the total energy deposited, we multiply $dE/dx$ in Eq. (\ref{dedx})
by the path length of the macro inside the human body, which we assume to be $\sim 10$ cm.}
 
{\bob Of course,} we are working with a very different range of projectile sizes and velocities from typical bullets.  Macros
have hypersonic velocities but very small geometric cross sections in our parameter
range of interest (as small as 1 micron$^2$).  Hence, their destructive effect is likely to be qualitatively different
from that of a bullet; a macro impact typically heats the cylinder of tissue carved out along its path
to a temperature of $10^7$ K \cite{p,q}, resulting
in an expanding cylinder of plasma inside the body.  While some studies have been done on collisions
of hypersonic, micron-sized projectiles with fixed targets \cite{u}, these differed significantly from
macro collisions in that the experimental projectiles were of much lower density, and the targets were ``hard" rather than
``soft."  Nonetheless, it is reasonable to take the kinetic energy deposited by a macro as a threshold
for significant damage to the human body.  Energy conservation requires that the macro energy ultimately
be deposited in the body in some form, whether mechanical or thermal, which will result in an equivalent amount of damage.
{\bob If anything, the unusual form of damage caused by a macro strike is likely to be {\it more} obvious and
easily detected than that of a bullet wound.}

{\bob We now perform a more detailed calculation.  We first require that the energy loss of the macros
in traversing the atmosphere be negligible, so that the macros reach their targets on the ground with
undiminished velocity.  We find that this corresponds to the bound
$\sigma_X/M_X \sim 10^{-4}$ cm$^2$ g$^{-1}$.  Macros above this threshold lose a significant amount of their
energy in the atmosphere and are therefore unconstrained by the argument considered here.  This consideration produces
the
diagonal upper bound on the blue excluded region in Figure 2.  Limiting our discussion
to macros satisfying this bound allows us to neglect
shielding from buildings, automobiles, or similar objects, since the column density of the atmosphere
is much larger than the column density of most inhabited structures.}

To determine the minimum macro cross section needed to cause significant human injury,
we assume macros possess a Maxwellian velocity distribution
\begin{equation}
	\label{eq:maxwellian}
	f_{MB}(v_X) = 
		\left( \frac{1}{\pi v_{vir}^2}\right)^{\frac{3}{2}}
		4\pi v_X^2 e^{-\left(\frac{v_X}{v_{vir}}\right)^2}, 
\end{equation}
where $v_{vir} \approx 250~ \,$km$\,$s$^{-1}$.  
This distribution is slightly modified by the motion of the Earth \cite{v}. We have also truncated this
distribution at the galactic escape velocity at the position
of the Solar System in the galaxy $v_{esc} \sim 550\,$km s$^{-1}$. 
Taking into account the distribution \eqref{eq:maxwellian},
multiplying Eq. (\ref{dedx}) by a path length of 10 cm and requiring the total energy deposited
to be of order 100 J or greater, we obtain the lower bound on $\sigma_X$ in our excluded region shown
in Fig. 2. This lower bound on the excluded region varies slowly with $M_X$ but is
roughly $\sigma_X \sim 10^{-7}$ cm$^2$.  Macros with cross sections below this bound would deposit
less than 100 J per human impact, so their interactions with human bodies 
{\color{black} might} not be noticeable.

{\bob For macros that have large enough cross sections to cause serious human injury or death, the rate of injuries
is proportion to the macro number density.  If we assume that the macros constitute the dark matter,
the total macro energy density is fixed at $\rho_{DM} \approx 5 \times 10^{-19}\,$g$\,$m$^{-3}$ \cite{w},
and the macro number density is inversely proportional to the macro mass:  $n_X = \rho_{DM}/M_X$.
Thus, the number of macro-human interaction events scales inversely as $M_X$, and our excluded region
will extend out to some upper bound on $M_X$.  To determine this upper bound, we argue
that there have been no unexplained injuries or deaths characteristic of macro collisions
among the well-monitored
population of the Western countries.}
{\color{black} Although there are many sudden unexpected deaths daily,
and a small fraction of these cannot be explained even following autopsy,
a death due to a macro strike would produce a striking signature, most likely a cylinder of vaporized
tissue surrounded by a larger cavity, with no projectile in evidence. }
{\glenn If the death occurred indoors, 
there would also have been associated damage 
to the structure, furnishings, {\it etc}.}
{\bob We assume that any such deaths would have been easily detected and well reported.
Hence, it is reasonable to take the observed number of such deaths in Western countries
over the past 10 years to be zero {\color{black}\footnote{{\color{black}At the low-energy (low cross-section) end of our constrained region, the destruction from a macro
would be similar to a gunshot, as we have noted in our paper.  Note that
deaths in this manner are always investigated by the authorities, with autopsies performed.
Furthermore, forensic pathologists go to great lengths to rule out other causes of
death, and occasionally discover that what appears to be a gunshot death is, in fact, due
to an entirely different cause (see, for example references
\cite{w1,w2}).  It is unlikely that a macro injury of this type would not be noted and reported upon
autopsy.  

The collision with a larger macro would likely produce a much more
destructive event.  It would be similar, in sheer destructive ability, to a meteor strike.  However, it
is believed with a high degree of confidence that no one, in modern times, has been killed by a meteorite
(see, e.g., references \cite{t1,w3}).
Given that the impact
of a large macro would be even more striking than a meteorite and leave even more unambiguous evidence,
we feel confident that such an event can be excluded.}}.}}
The expected number of macro passages through a population of $N$ humans 
depends on $M_X$ as
\begin{equation}\label{Nevents}
N_{events}=f\frac{\rho_{DM}N A_{human}T_e v_X}{M_X} 
	\,,
\end{equation}
where 
$A_{human} \sim 1\,$m$^2$ is the typical cross-sectional area of a human,
$N \approx 8\times 10^8$ is the population of the US, Western Europe and Canada
and $T_e$ is the exposure time, 
which we take to be 10 years.
We also take into account the distribution \eqref{eq:maxwellian}
by adding in an additional factor f accounting for
the fraction of macros in the distribution that possess a
minimum velocity.
Considering the entire distribution, we find that
\begin{equation}\label{Nevents}
N_{events}\approx \frac{80000 g}{M_X} 
	\,,
\end{equation}

Since the impact of a macro on a human
is a Poisson process,
the probability $P(n)$ of 
$n$ impacts over the exposure time $T_e$
follows the Poisson distribution:
\begin{equation}
\label{eq:Poisson}
P(n) = \frac{{N_{events}}^n}{n!}
e^{N_{events}}\,.
\end{equation}
\newline
where $N_{events}$ is the expected
number of events per interval,
as calculated in equation \eqref{Nevents}.
Having observed no macro-related deaths {\color{black}or serious injuries} over 10 years in this population,
we may constrain
$M_x \leq {\bob 5 \times 10^4}\,$g at the 95$\%$
level; 
{\bob this is the vertical line demarcating the right-hand
boundary of the excluded region in Figure 2.  To be conservative, we have also considered the possibility
of a nonzero number of macro deaths.  The corresponding limits are displayed in Figure 2 for 10 deaths over the past
10 years (medium red).
We see that even in this case we can exclude a significant region of parameter space not currently constrained
by the geological (mica) limits.}

{\bob Our excluded region, then, is the roughly triangular region shown in Fig. 2.  While it does have
some overlap with the mica constraint, it also excludes a wide range of previously-allowed parameter space.
One might hope that stronger limits could be derived, e.g., from domestic livestock or wild animals, but this is not the
case.  The biomass of livestock is roughly twice the biomass of humans \cite{x}, but the deaths
of domestic animals are considerably less well-monitored.  Furthermore,
livestock and humans together outmass all wild vertebrates
combined, with the exception of fish \cite{x}, so it is unlikely that useful constraints could be
derived from the deaths of wild animals.}

\section{Conclusion}
We have considered a phenomenological approach
and constrained the abundance of macros over the relevant mass range
based on the null observation of unique human injuries/deaths.
{\bob  These new limits complement the searches proposed in Refs. \cite{p,q}}.
The results presented here constrain macros
with physical sizes as small as
several microns and masses less than {\bob $50\,$kg (Fig. 2); these are}
significantly smaller than the cross-sections
that we expect to probe in \cite{p,q}
over the same mass range. However,
the methods outlined in \cite{p,q}
will be able to probe larger masses than {\bob those}
constrained in this manuscript.
Admittedly, the effect of impacts of {\bob hypersonic objects} smaller than {\bob a few} microns
on {\bob the human body} remains an open question, so
{\bob more detailed} {\glenn analysis} {\bob might allow constraints on even smaller macro cross sections.
Regardless,} our results open a new window on dark matter:  the human
body as a dark matter detector.

\includegraphics[width=\textwidth]{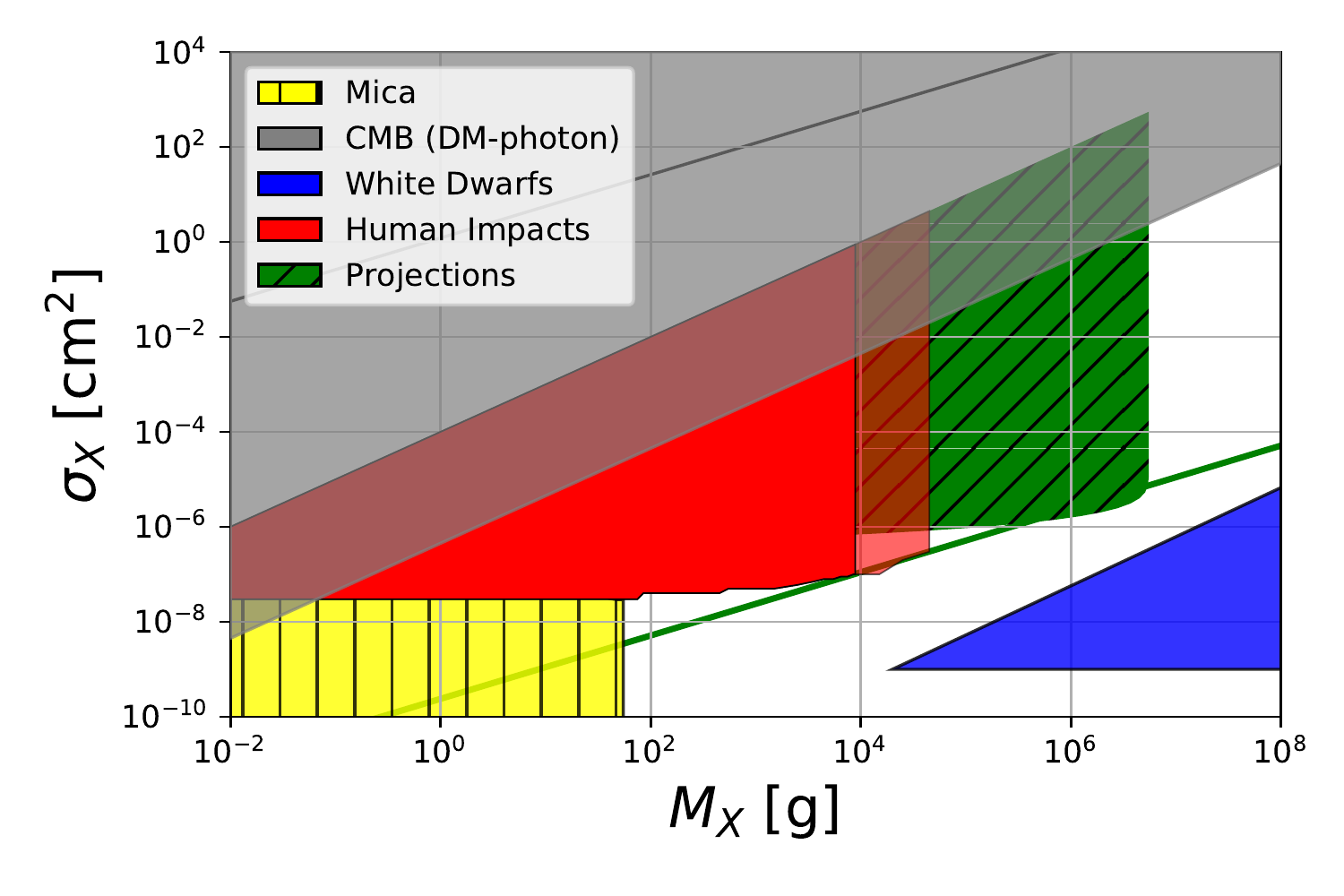}
\captionof{figure}{
Constraints in yellow are
derived from a lack of tracks in an
ancient slab of mica \cite{h,i}, in grey 
from the Planck Cosmic Microwave Background data considering
elastic macro-photon interactions \cite{j},
in red from a lack of human impacts (this work) and in blue
from thermonuclear runaway in white dwarfs \cite{n}.
The red excluded region is based on 
fewer than 10 macro deaths (medium red),
and zero macro deaths (light red) over the past 10 years
in the population of the US, Canada, and Western Europe.
{\color{black} The green hatched region
represents projections from other proposed
ways of probing macro parameter space \cite{p,q}.}}

\section*{Acknowledgements} This work was partially supported by Department of
Energy grant de-sc0009946 to the particle astrophysics
theory group at CWRU. 
R.J.S. was partially supported by the  Department of Energy,
de-sc0019207.  R.J.S. thanks M.S. Hutson for helpful discussions.

\end{document}